\documentclass[showpacs,amsmath,amssymb,pra,superscriptaddress,floatfix,twocolumn]{revtex4-1}

\usepackage{graphicx}
\usepackage{color}
\usepackage{amsmath,bm}
\usepackage{amssymb}
\usepackage{hyperref}
\usepackage{float}
\definecolor{darkblue}{rgb}{0.00,0.00,0.55}
\definecolor{darkgreen}{rgb}{0.00,0.39,0.00}

\newcommand{\ry}{Rydberg }

\newcommand{\ie}{i.\,e.}%

\def\braket#1{\mathinner{\langle{#1}\rangle}}

\begin{document}
\title{Ultralong-Range Rb-KRb Rydberg Molecules: Selected Aspects
of Electronic Structure, Orientation and Alignment}

\author{Javier Aguilera-Fern\'andez}
\affiliation{Instituto Carlos I de F\'{\i}sica Te\'orica y Computacional and
      Departamento de F\'{\i}sica At\'omica, Molecular y Nuclear, Universidad de Granada, 18071
      Granada, Spain}

\author{H. R. Sadeghpour}
\affiliation{ITAMP, Harvard-Smithsonian Center for Astrophysics, Cambridge, Massachusetts 02138, USA} 
  
\author{Peter Schmelcher}
\affiliation{The Hamburg Center for Ultrafast Imaging, Luruper Chaussee 149, 22761 Hamburg, Germany}
\affiliation{Zentrum f\"ur Optische Quantentechnologien, Universit\"at Hamburg,
Luruper Chaussee 149, 22761 Hamburg, Germany} 

\author{Rosario Gonz\'alez-F\'erez}
\affiliation{Instituto Carlos I de F\'{\i}sica Te\'orica y Computacional and
      Departamento de F\'{\i}sica At\'omica, Molecular y Nuclear, Universidad de Granada, 18071
      Granada, Spain}

\begin{abstract}
We investigate the structure and features  of an ultralong-range triatomic Rydberg molecule formed by a
Rb  Rydberg atom and a KRb diatomic molecule. 
In our numerical description, we perform a realistic treatment of the internal rotational motion of the 
diatomic molecule,
and take into account the Rb($n, l\ge 3$) \ry degenerate manifold and the energetically closest neighboring levels
with  principal quantum numbers $n'>n$ and orbital quantum number $l\le2$.
We focus here on the adiabatic  electronic  potentials evolving from the Rb($n, l\ge 3$) and Rb($n=26, l=2$)
 manifolds. 
The directional properties of the KRb diatomic molecule within 
the Rb-KRb triatomic \ry molecule are also analyzed in detail. 
\end{abstract}

\maketitle

\section{Introduction}

Recently the existence of  a new class of ultralong-range polyatomic  \ry molecules was predicted
theoretically~\cite{rittenhouse10,rittenhouse11}.  These huge-sized
\ry molecules  are  formed by  a \ry atom and  either  a $\Lambda$-doublet 
or  a rotating polar diatomic molecule~\cite{rittenhouse10,Rosario}.
The binding mechanism is established by the electric field of the \ry electron and the
ionic core that couples either the two internal states or hybridizes the rotational motion
of the molecule due to its permanent electric dipole moment.
The existence of these ultralong-range triatomic \ry molecules was predicted for polar diatomic molecules with 
subcritical electric dipole moments ($d_0 < 1.639$~D) in order to prevent the binding of the
 \ry electron to the heteronuclear (polar) diatomic molecule~\cite{fermi47,turner77,clark79,fabrikant04}.

These giant \ry  molecules  possess an electronic structure characterized by 
oscillating Born-Oppenheimer potential curves evolving from the  Rb($n, l\ge 3$) \ry manifold with binding  energies of a few GHz. Their electronic structure could 
be controlled and  manipulated by weak electric fields of a few V/cm~\cite{mayle12}.
The Rydberg-field-induced coupling  gives rise to a strong hybridization of the angular motion
of the diatomic molecule together with a strong orientation and alignment.   
In particular, the orientation of the diatomic molecule within the \ry molecule changes sign and strongly depends
on the dominant contribution to the \ry electric field~\cite{rittenhouse10,Rosario}. 
The two internal rotational states of opposite orientation of the polar diatomic molecule 
could be coupled by means of a Raman scheme~\cite{rittenhouse10}, which 
might allow for the switchable dipole-dipole interaction needed to implement molecular
qubits~\cite{kuznetsova}.

Recently, we investigated electronic properties of the polyatomic Rydberg molecule
composed of a Rb
 \ry atom and the KRb diatomic molecule~\cite{Rosario}. 
Our  theoretical description has included an explicit 
treatment of the angular degrees of freedom of the diatomic molecule within the rigid-rotor approximation.
Here, we extend this study on the Rb-KRb  triatomic \ry molecule.  
We investigate the molecule and obtain Born-Oppenheimer potential (BOP) curves of
 the  Rb($n, l\ge 3$)-KRb Rydberg molecule with varying principal quantum number $n$ 
 of the \ry electron and varying internuclear distance between the ionic  core Rb$^+$ and the diatomic KRb.
 The lowest-lying BOPs  evolving from  Rb($n, l\ge 3$) and KRb($N=0$) present potential wells with depths of a few GHz.
We also explore the adiabatic potentials  of the \ry molecule formed by the KRb being in 
 a rotational excited state, and the Rb($26d$) \ry  manifold. These BOPs evolving from the  Rb($26d$) state 
 present wells
 with depths of a few MHz that  support several vibrational bound states. 
This opens the possibility of creating these macroscopic \ry molecules by two-photon excitation of ground-state 
Rb in an ultracold mixture of Rb and KRb.
The orientation and alignment of the diatomic molecule within the \ry molecule is also analyzed in terms of
the contributions to  the electric field due to the \ry electron and ionic core. 

The paper is organized  as follows. In~\autoref{sec:hamiltonian} we describe the adiabatic Hamiltonian of the 
triatomic \ry molecule. 
The electronic structure of a Rb-KRb  \ry   molecule is analyzed in detail 
in~\autoref{sec:results}: We analyze the Born-Oppenheimer potentials evolving 
 from the degenerate \ry  manifolds Rb($n, l\ge 3$) and Rb($26d$), and 
 the directional properties of the KRb diatomic molecule within
the  Rb-KRb \ry trimer. The conclusions are provided in~\autoref{sec:conclusions}.
\section{The adiabatic molecular Hamiltonian}
\label{sec:hamiltonian}

We consider a triatomic \ry molecule formed by a Rydberg atom and a diatomic 
heteronuclear molecule, which is located on the $Z$ axis 
in the laboratory fixed frame (LFF) at a distance $R$ from the ionic core, 
the latter being placed at the origin of the LFF. The Born-Oppenheimer Hamiltonian reads
\begin{equation}
H_{ad}=H_{A}+H_{mol}
\label{sys_halmilt}
\end{equation}
where $H_{A}$ is the single electron Hamiltonian describing the Rydberg atom
\begin{equation}
H_{A}=-\dfrac{\hslash^{2}}{2m_{e}}\triangledown^{2}_{r}+V_{l}(r),
\end{equation}
with $V_{l}(r)$ being the $l$-dependent model potential~\cite{Marinescu},  
and  $l$ the angular quantum number of the Rydberg electron. 

The term $H_{mol}$ is the Hamiltonian of the rotational motion of the diatomic molecule
in the electric field due to the Rydberg electron and ionic core
\begin{equation}
H_{mol}=B\textbf{N}^{2}-\textbf{d}\cdot \textbf{F}_{ryd}(\textbf{R},\textbf{r})
\label{int_hamilt}
\end{equation}
with $\textbf{N}$ being the molecular angular momentum operator, $B$  the rotational constant,  and $\textbf{d}$  the 
permanent electric dipole moment. Note that we describe the rotational motion of the diatomic molecule within 
the rigid-rotor approximation.
The electric field created by the ionic core and Rydberg electron at position $\mathbf{R}$ reads~\cite{rittenhouse10,rittenhouse11}
\begin{equation}
F_{ryd}(\mathbf{R},\mathbf{r})=e\dfrac{\mathbf{R}}{R^{3}}+e\dfrac{\mathbf{r}-\mathbf{R}}{|\mathbf{r}-\mathbf{R}|^{3}}
\label{int_ryd}
\end{equation}
with $\mathbf{R}=R\hat{Z}$. We consider all three spatial components of this electric field taking into 
account that the position of the polar molecule is fixed along the LFF $Z$-axis. 
For a detailed expression of this electric field, we refer the reader to Refs.~\cite{Rosario, ayuel09}.  


The orbital angular momentum of the \ry electron and the rotational angular momentum of the diatomic
molecule are coupled to
the total angular momentum  of the \ry molecule (excluding an overall rotation) 
$\mathbf{J}=\mathbf{N}+\mathbf{l}$. For the considered configuration of the \ry molecule, the 
eigenstates are characterized by the magnetic quantum number $M_J$, which is 
 the  projection of $\mathbf{J}$ onto the LFF $Z$-axis.

The Schr\"odinger equation associated with the Hamiltonian \ref{sys_halmilt}  is solved by 
a basis set expansion in the coupled basis 
\begin{eqnarray}
\psi_{nlNJM_{J}}(\mathbf{r},\Omega_d)&=&\sum^{m_{l}=l}_{m_{l}=-l}\sum^{M_{N}=N}_{M_{N}=-N} \braket{lm_{l}NM_{N}|JM_{J}}\nonumber\\
&&\times \psi_{nlm}(\textbf{r})\times Y_{NM_{N}}(\Omega_{d})
\end{eqnarray}
where $\psi_{nlm}(\textbf{r})$ is the \ry electron wave function, 
 $Y_{NM_{N}}(\Omega_{d})$ the spherical harmonics representing 
 the field-free rotational wave function of the diatomic molecule with   the rotational and magnetic quantum numbers
 $N$ and $M_{N}$, respectively,
 $\Omega_{d}$ the internal coordinates of the diatomic molecule, 
  and  $\braket{lm_{l}NM_{N}|JM_{J}}$ the Clebsch-Gordan coefficient. 
Taking into account the azimuthal symmetry of the \ry molecule, the basis set expansion of the wave function reads
\begin{equation}
\Psi(\mathbf{r},\Omega_d;R)=
\sum_{n,l,N,J}C_{nlNJ}(R)\psi_{nlNJM_{J}}(\mathbf{r},\Omega_d), 
\label{basis_expansion}
\end{equation}
where  we have explicitly indicated  the dependence of  the wave function and 
expansion coefficients $C_{nlNJ}(R)$  on the internuclear distance $R$ between the diatomic molecule and the ionic core.
For computational reasons, of course, we have to cut the infinite series \eqref{basis_expansion} to a finite one. 
In this work, we have included  the rotational excitations with 
$N\le  6$ for KRb,  and for Rb, the   ($n,l\ge3$) degenerate manifold, and the 
energetically neighboring levels $(n+1)d$, $(n+2)p$, and $(n+3)s$. 
The quantum defect of the  Rb($nf$) \ry state  has been neglected.

\section{The electronic structure of the Rb-KRb \ry molecule }
\label{sec:results}

In this section, we investigate the electronic structure of the  \ry molecule formed by a Rb atom and the heteronuclear 
molecule KRb  with  rotational constant $B=1.114$~GHz~\cite{ni09}, and  electric dipole moment 
$d=0.566$~D~\cite{ni08}.
The interaction between the electric field created by the ionic core and \ry electron and the permanent
 electric dipole moment of KRb is responsible for the binding mechanism of Rb-KRb.  
The electric field  \eqref{int_ryd} decreases as the distance between Rb$^+$ and KRb  is increased, and, 
therefore, the coupling between the subsystems Rb and KRb also decreases.
For large enough values of $R$, the  adiabatic electronic potentials 
approach the energies of the uncoupled system  $E_{nl}+BN(N+1)$,
with $E_{nl}$ being the energy of Rb($n,l)$,  and $BN(N+1)$ the rotational energy
of KRb in an eigenstate with rotational quantum number $N$.

\begin{figure*}
\centerline{\includegraphics[scale=0.6]{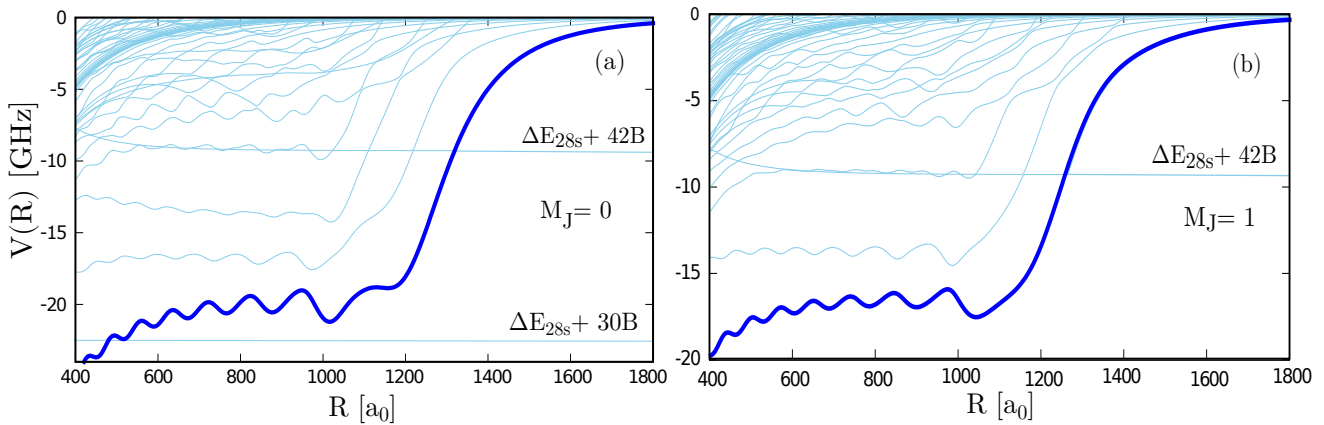}}
\caption{For the Rb-KRb \ry molecule, the Born-Oppenheimer potentials as a function of the distance between
KRb and  Rb$^+$  for (a) $M_{J}=0$ and  (b) $M_{J}=1$ are shown, specifically the BOP evolving from 
Rb($28s$) and KRb in a rotational excited state, and   from the Rb($n=25, l\ge 3$) manifold
and the rotational ground state of Rb. The  energetically lowest-lying adiabatic potentials evolving from
the Rb($n=25, l\ge 3$) manifold are shown with thicker lines.
The zero energy has been set to the energy of the Rb($n=25, l\ge 3$) degenerate manifold and the KRb
($N=0$) state.}
\label{espectro_25}
\end{figure*}

\subsection{The adiabatic potentials evolving from the Rb($n,l\ge3$)  degenerate manifold}

The BOPs  of the Rb-KRb molecule evolving from the \ry degenerate manifold 
Rb($n=25,l\ge3$)  and with magnetic quantum numbers $M_J=0$ and $M_J=1$ are presented 
in~\autoref{espectro_25}~(a) and~\autoref{espectro_25}~(b), respectively.  
In this region of the spectrum, we encounter  adiabatic states evolving from  Rb$(28s)$ and 
KRb in a rotationally excited state, which show a very weak dependence on the internuclear 
distance. This behavior is due to the small impact of the \ry electric field on the $N=6 $ excited state of KRb, which possesses a
large field-free rotational energy.
The electronic states evolving from the Rb($n=25,l\ge3$)  manifold are shifted in energy 
from the uncoupled-system 
energy  $E_{nl}+BN(N+1)$. These energy shifts strongly depend on the internuclear separation 
of Rb-KRb due to the interplay between the electric fields  created by the \ry electron and  the ionic core. 
They are of the range from a few to tens of GHz. 
The BOPs oscillate as a function of  $R$, and numerous avoided crossings among neighboring 
electronic states characterize the spectrum.
For $R\sim1000~a_0$, the energetically lowest-lying BOPs present potential wells of a few GHz depths, 
which support several vibrational states~\cite{Rosario}. 
By further increasing $R$, the coupling between the \ry atom and KRb decreases, and the  BOPs 
approach  the asymptotic limit   of the uncoupled system. Note that in these figures, the zero energy has been set at 
$E_{n=25,l\ge3}$, \ie, for the uncoupled system KRb($N=0$) and Rb($n=25, l\ge 3$).

\begin{figure*}
\centering
\includegraphics[scale=0.86,angle=0]{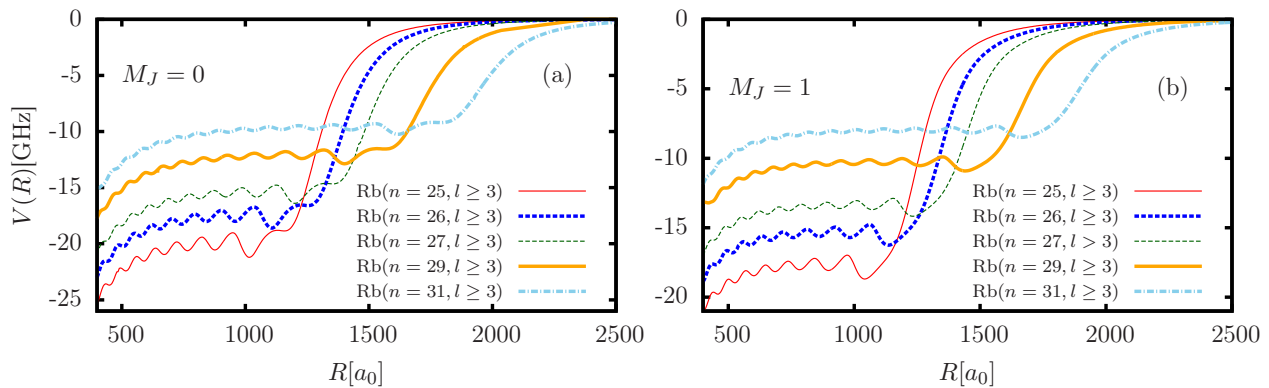}
\caption{For the Rb-KRb molecule, the
energetically lowest-lying states with $M_J = 0$ and $M_J =1$ evolving from
the Rb($n, l\ge 3$) \ry manifold, for $n=25, 26, 27, 29$,  and $31$, and KRb in the rotational 
ground state $N=0$ are shown. To facilitate the comparison, the zero 
energy has been set to the energy of the Rb($n, l\ge 3$) degenerate manifold and the KRb($N= 0$) level.}
\label{BOP-n-states}
\end{figure*}
We focus  now on the evolution of the electronic structure as the principal quantum number of the \ry
degenerate manifold Rb($n, l\ge3$) is increased. 
In~\autoref {BOP-n-states}, we present 
the energetically lowest-lying adiabatic electronic states with $M_J=0$ and á$M_J=1$ evolving 
from the  Rb($n, l\ge 3$)  manifolds for  $n=25, 26, 27, 29$, and  $31$.
These BOPs show a qualitatively similar oscillatory behavior as a function of $R$.
By increasing the  principal quantum number of the 
\ry manifold, some  differences are observed. 
First, the energy shift of these BOPs from the degenerate manifold of the \ry atom decreases as $n$ is increased. 
Second, the location of the potential wells is shifted towards larger values of the 
internuclear separation between Rb$^+$  and KRb. This is due to the radial
 extension of the electronic \ry wave 
function of  Rb($n, l\ge 3$) which increases as $n$ is enhanced. 
Third, the depth of the wells decreases with increasing $n$.
The  outermost  potential well of the $M_J=0$ BOPs is very shallow and do not support vibrational bound states. In 
contrast, the potential well  with $M_J=0$  located at the left of this  outermost one is the deepest one 
supporting several bound states and with depths of several GHz for all the Rb($n, l\ge 3$)-KRb($N=0$) molecules investigated here. 
For instance,  we have found $7$ and $6$ vibrational bound states in these potential wells
 of these  energetically lowest-lying  Rb($n=25, l\ge 3$) and  Rb($n=31, l\ge 3$)  BOPs with $M_J=0$, respectively.

\begin{figure*}
\centering
\includegraphics[scale=0.85,angle=0]{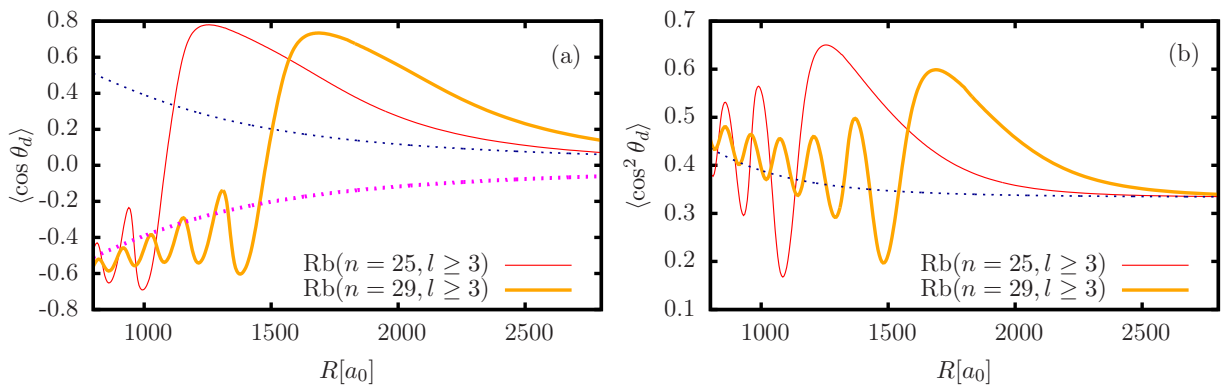}
\caption{
For the KRb  molecule within the energetically lowest-lying BOPs  with $M_J = 0$ of
Rb($n=25, l\ge 3$)-KRb  (thin solid) and Rb($n=29, l\ge 3$)-KRb (thick solid),  
we show (a) the orientation $\langle\cos \theta_{d} \rangle$  and (b) the alignment   
$\langle\cos^2 \theta_{d} \rangle$  along the LFF $Z$-axis
 versus the internuclear distance between Rb$^+$ and KRb. 
We also present the orientation and alignment of  KRb in an  electric field parallel  (thick dotted) and  antiparallel  (thin dotted) to the LFF $Z$-axis and  strength $5.14/R^{2}$~GV/cm.}
\label{BOP-n-orientation}
\end{figure*}
\begin{figure}
\includegraphics[scale=0.82,angle=0]{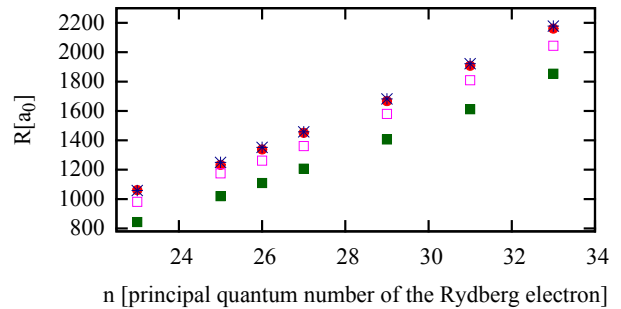}
\caption{
For the Rb-KRb  \ry molecule, 
we show the distance between the Rb$^+$ core and KRb   (\textcolor{red}{$\bullet$})
at which the orientation of KRb is maximal within the lowest-lying $M_J=0$ adiabatic potentials evolving from the
Rb($n, l\ge 3$)  manifolds for varying principal quantum number $n$.  For these lowest-lying  BOPs, we also present the internuclear 
separation $R$ of Rb-KRb at which the 
one before the outermost (\textcolor{magenta}{$\square$}) 
and the outermost (\textcolor{darkgreen}{$\blacksquare$}) potential wells reach the minimum value. 
For comparison, we 
show the classical turning point $R\sim 2n^2 a_0$
(\textcolor{darkblue}{$\star$}).}
\label{max-orientation}
\end{figure}

We analyze now the directional features of the KRb molecule within the Rb-KRb triatomic \ry molecule.
For the lowest-lying electronic  potentials evolving from the 
Rb($n=25, l\ge 3$) and Rb($n=29, l\ge 3$) manifolds and with $M_J=0$,   
the orientation, $\langle\cos \theta_{d} \rangle$, 
and alignment, $\langle\cos^2 \theta_{d} \rangle$, 
of KRb along 
the LFF $Z$-axis are shown in~\autoref{BOP-n-orientation}~(a) and~\autoref{BOP-n-orientation}~(b), respectively. 
For the sake of simplicity, we only analyze the results  for the Rb-KRb in these two manifolds, but  
qualitatively similar behavior is obtained for the others BOPs presented  in~\autoref{BOP-n-states}.
We have computed the orientation and alignment  of the rotational ground state of  KRb  in 
an electric field  
of varying field strength $5.14/R^{2}$~GV/cm, \ie, the strength of the 
electric field due to the  \ry core, being parallel and  antiparallel to the LFF $Z$-axis,  see~\autoref{BOP-n-states}. 
These results allow us to get a deeper physical insight 
into the interplay between  the electric field due to the \ry electron and the Rb$^+$ core. 
For both BOPs, we encounter a regime where the electric field due to the \ry core is dominant, \ie,
the total electric field is antiparallel to the LFF $Z$-axis, and KRb is anti-oriented. 
In this regime,  the orientation of  KRb within Rb-KRb is very close to the orientation of 
KRb  in the $5.14/R^{2}$~GV/cm electric field. 
Let us comment, that for $R=1000~a_0$, the electric field strength due to the \ry core is $5$~kV/cm. 
The oscillations of the orientation of  the KRb within Rb-KRb
are due to the \ry electrons electric field, which oscillates resembling the behavior of its wave function. 
The amplitude of these oscillations increases as the  \ry electrons electric field becomes more important. 
When the contribution of this field becomes dominant, the orientation of KRb changes sign
and  KRb becomes oriented  along the $Z$-axis.
The differences in the  electronic \ry wave  function of Rb($n=29, l\ge 3$) and Rb($n=25, l\ge 3$), are reflected in 
the fact that the KRb molecule is anti-oriented for a larger  range of values of $R$ for 
Rb($n=29, l\ge 3$)-KRb($N=0$).
KRb is maximally oriented around the classical turning point $R\sim 2n^2a_0$, and we have
$\langle\cos \theta_{d} \rangle\approx 0.8$, which is the orientation of KRb in a strong electric field of
$50$~kV/cm strength~\cite{Sanchez}.
The internuclear separation of Rb-KRb at which the KRb orientation reaches the maximum  is shown versus the 
principal quantum number  of the  Rb($n, l\ge 3$) degenerate 
manifold  in~\autoref{max-orientation}.
 We observe a very good agreement with the $2n^2a_0$ behavior. 
In ~\autoref{max-orientation} we show also  the internuclear separation of Rb-KRb at which the 
energetically lowest-lying BOP with $M_J=0$ reaches the two outermost minima.  The radial position of these two 
minima is shifted to lower values of $R$ compared to the maximum of the KRb orientation. 
By further increasing $R$, the orientation 
 decreases towards zero and approaches the orientation of KRb in the  $5.14/R^{2}$~GV/cm electric field. 
 Note that in the absence of an electric field, the KRb molecule is not  oriented. 

The alignment along the LFF $Z$-axis of KRb within Rb-KRb also  shows an oscillatory 
behavior around the alignment of  the KRb  in a  $5.14/R^{2}$~GV/cm 
electric field, cf.~\autoref{BOP-n-orientation}~(b). 
As the coupling of the dipole moment of KRb with the  electric field due to the \ry electron becomes dominant compared to the coupling with the Rb$^+$ electric field,   the amplitude of the 
oscillations become very large. When the distance of KRb form the Rb$^+$ core is 
large, the alignment approaches the field-free value  $\langle\cos^2 \theta_{d} \rangle\approx 0.3333$.

\subsection{The adiabatic potentials evolving from the Rb($26d$) \ry state}
\begin{figure*}
\centering
\includegraphics[scale=0.86,angle=0]{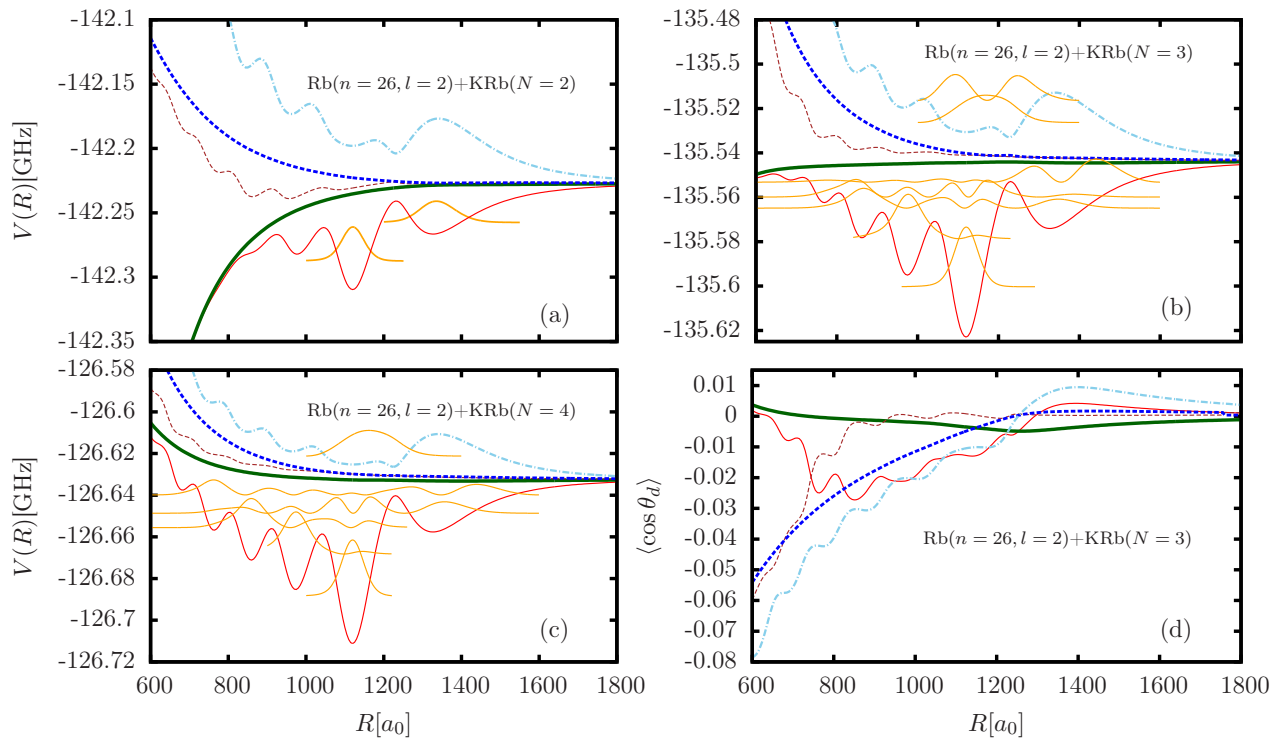}
\caption{For the Rb-KRb \ry molecule, 
Born-Oppenheimer potentials  with $M_J=0$ evolving from the Rb($26d$) state and KRb in the rotational levels  (a)
 $N=2$,  (b)  $N=3$, and (c) $N=4$. We have plotted the 5 electronic  states with $M_J=0$ and  total angular momentum $J=2-N,\dots N+2$ in the uncoupled system.
In the potentials wells, the square of the vibrational wave functions are shown in arbitrary units.
The zero energy has been set to the energy of the Rb($n=25, l\ge 3$) degenerate manifold and the KRb
($N=0$) state.
In panel (d), we show the orientation along the LFF $Z$-axis of KRb within the BOPs of the \ry molecule
Rb($26d$)-KRb($N=3$). }
\label{d-states}
\end{figure*}

In the absence of an external electric field, the ultralong-range \ry molecule could 
be created from an ultracold  mixture of  Rb  and KRb, by exciting the Rb from the 
 ground-state to a $nd$ \ry level in  a two-photon process. In this section, we investigate 
 the BOPs of the \ry molecule formed from Rb($26d$).

The adiabatic potentials with $M_J=0$ evolving from the Rb($26d$) \ry states and 
the diatomic molecule in the rotational excitations $N=2$, $3$ and $4$ are present in~\autoref{d-states}.   There are 5 electronic  states with $M_J=0$ and  total angular momentum $J=2-N,\dots N+2$ in the uncoupled system.
These BOPs  show either a strong or a weak dependence on $R$, and 
when $R$ is increased beyond a certain value they approach the energetical limit $\Delta E_{26d}+ BN(N+1)$, 
with  $N=2$, $3$ and $4$ and $\Delta E_{26d}=|E_{24,3}-E_{26d}|$. 
The  BOPs with oscillatory behavior possess potential wells having depths of a few MHz or tens of MHz. 
These potential wells are deep enough to accommodate at least one vibrational level in which the 
polyatomic \ry  molecule would exist. 
For a few vibrational levels, the square roots of the wave functions are shown in the potential wells 
in~\autoref{d-states}. 
For the BOPs evolving from the  $N=3$ and $N=4$ rotational states of KRb, we encounter several vibrational 
bound states  with wave functions extending over several potential wells.

In~\autoref{d-states}~(d), we present the orientation of the diatomic molecule  along the LFF $Z$-axis for
the electronic potentials evolving from the Rb($26d$) \ry manifold and KRb ($N=3$). 
The diatomic molecule is weakly oriented  along the LFF $Z$-axis, and similar results are
obtained for the BOPs evolving from the diatomic molecule in the rotational excitations $N=2$ and $N=4$. 
This weak orientation can be explained in terms of the small impact of the \ry  electric field 
on the rotational excited state,  which is  smaller than for the rotational ground state due to its larger
rotational energy~\cite{Sanchez}. 
Thus, the  weaker the coupling between the $N$ rotational state and the neighboring levels $N\pm 1$,
 the smaller the orientation is.

\section{Conclusions}
\label{sec:conclusions}

In this work, we have considered the  ultralong-range  \ry molecule formed by the KRb diatomic 
molecule and the \ry Rb atom.  
The coupling mechanism of this triatomic \ry  molecule is based on the interaction between the electric field 
produced by the \ry electron and core and the permanent electric dipole moment of the polar molecule. 
We have performed an analysis of the electronic structure of Rb-KRb when the
Rb atom is in  the \ry degenerate manifold ($n, l\ge 3$), with $n=25, 26, 27, 29$,  and $31$. 
These adiabatic potentials show an oscillatory behavior as a function of the 
separation between  Rb$^+$ and KRb. For the Rb($n, l\ge 3$)-KRb($N=0$) molecules, the lowest-lying BOPs  show potential wells of a few GHz depths.  
As the principal quantum number of the \ry excitation $n$ is enhanced, the
depths of these potential wells decrease and their positions are  shifted towards larger values of the internuclear 
separation $R$. These features reflect the oscillatory structure of the Rydberg electronic wave function. 

We have also investigated the properties of the \ry molecule formed by the Rb($26d$) \ry state
and the diatomic molecule in an excited rotational state. BOPs with both an oscillatory  
and a smooth behavior  as function of $R$ are encountered.
For   Rb($26d$)-KRb(N), these oscillatory  BOPs  present potential wells  with depths of a few tens 
of MHz, which can accommodate a number of bound vibrational states. 
These results demonstrate that  these ultralong-range \ry molecules could be created by a standard 
two-photon excitation of ground state Rb in an ultracold mixture of Rb and KRb. 

Due to the electric field of the \ry electron and core, the directional properties of the KRb molecule 
within the Rb-KRb triatomic molecule are  significantly affected.
The orientation and alignment of KRb  has been analyzed in terms of the contribution of the electric fields due to 
the \ry electron and ionic core.
 Within these BOPs, 
 the polar diatomic molecule  presents two opposite orientations along the LFF $Z$-axis
 with its electric dipole moment pointing either toward or away from the ionic core. 

\begin{acknowledgments}

RGF and JAF 
gratefully acknowledge financial support by the Spanish projects FIS2014-54497-P (MINECO),
P11-FQM-7276 (Junta de Andaluc\'{\i}a), and by the Andalusian research group FQM-207.
We also acknowledge financial support by the Initial Training Network COHERENCE of the European Union
FP7 framework. 
HRS and PS acknowledge ITAMP at the Harvard-Smithsonian Center for Astrophysics for
support. 
\end{acknowledgments}



\end{document}